# Pressure-induced superconductivity in kagome single crystal $Pd_3P_2S_8$

Ying Zhou[1], Xinyi He[1], Shuyang Wang[2,3], Jing Wang[2,3], Xuliang Chen[2,3], Yonghui Zhou[2,3], Chao An[1], Min Zhang[1], Zhitao Zhang[2,3,*] and Zhaorong Yang[1,2,3,*]

[1]*Institutes of Physical Science and Information Technology, Anhui University, Hefei 230601, China*
[2]*Anhui Province Key Laboratory of Condensed Matter Physics at Extreme Conditions, High Magnetic Field Laboratory, HFIPS, Chinese Academy of Sciences, Hefei 230031, China*
[3]*High Magnetic Field Laboratory of Anhui Province, Hefei 230031, China*

[*]Corresponding authors: ztzhang@hmfl.ac.cn and zryang@issp.ac.cn.

Kagome lattice offers unique opportunities for the exploration of unusual quantum states of correlated electrons. Here, we report on the observation of superconductivity in a kagome single crystal $Pd_3P_2S_8$ when a semiconducting to metallic transition is driven by pressure. High-pressure resistance measurements show that the metallization and superconductivity are simultaneously observed at about 11 GPa. With increasing pressure, the superconducting critical temperature $T_c$ is monotonously enhanced from 2.6 K to a maximum 7.7 K at ~52 GPa. Interestingly, superconductivity retains when the pressure is fully released. Synchrotron XRD and Raman experiments consistently evidence that the emergence of superconductivity is accompanied with an amorphization and the retainability of superconductivity upon decompression can be attributed to the irreversibility of the amorphization.

**I. Introduction**

Generally, the geometrically frustrated kagome lattice is expected to have both flat band and Dirac dispersion bands, where the former can promote the electron correlations through destructive interference induced localization [1,2]. With the presence of the peculiar electronic structure and strong correlation effects, kagome materials often exhibit various and highly tunable electronic instabilities, including the spin liquid state, anomalous Hall effect, unconventional superconductivity, and so on [3-8]. Particularly, flat band systems are considered as ideal platforms to explore for exotic superconductivity [9-14]. For example, theoretical studies proposed that high-$T_c$ superconductivity could be realized in flat band systems through utilizing the inter-band pair scatterings between dispersion bands and the flat band which has divergent density of states [9-11].

$Pd_3P_2S_8$ is a two-dimensional kagome semiconductor which so far is mainly utilized as nanomaterials for photodetection, sensing, catalyzing, etc [15-19]. Recently, DFT calculations demonstrated that $Pd_3P_2S_8$ has a flat band locating right below the Fermi level which is primarily from Pd 4$d$-orbitals [20,21]. Therefore, it would be of particular interest to explore for superconductivity in $Pd_3P_2S_8$ if a metallization transition can be induced, for example, by element doping or pressure. Yan *et. al.* showed that Se substitution for S atoms is able to reduce the bandgap to some extend [20]. However, the solubility limit of Se substitution (<25%) prevents a potentially existed metallization from taking place [20].

In this paper, we grew high-quality $Pd_3P_2S_8$ single crystals and performed comprehensive high-pressure electrical resistance, Raman and XRD measurements to investigate the pressure-evolution of electronic properties and to search for superconductivity in the kagome system $Pd_3P_2S_8$. At a critical pressure of about 11 GPa, we observe concurrence of metallization and superconductivity with $T_c$ of 2.6 K. The superconductivity is continuously enhanced by further compression and survives after the pressure is reduced to ambient pressure. Based on structural analyses, we find that the occurrence of superconductivity and its retainability upon decompression are directly related to an irreversible amorphization process.

## II. Experimental Methods

$Pd_3P_2S_8$ single crystals were synthesized via chemical vapor transport method with iodine as a transport agent [21]. The phase purity of the samples was determined by x-ray diffraction (XRD) measurements on single crystal and powdered crystals. Stoichiometry of the crystals was confirmed by energy dispersive x-ray spectrometry (EDXS) with area and point-scanning modes. Optical transmittance absorption spectra were collected by using a UV/Vis/NIR spectrometer. High-pressure electrical transport experiments were performed in a nonmagnetic Be-Cu diamond-anvil cell using an in-house 9-T transport property measurement system [22]. High-pressure Raman scattering and angle-dispersive synchrotron XRD measurements were conducted in Mao-Bell cell with Daphne 7373 as transmitting medium. The Raman spectra were recorded with cleaved single-crystal flakes using a commercial Renishaw spectroscopy system with a 532 nm laser excitation line. Synchrotron XRD experiments were performed with powdered crystals at the beamline BL15U1 of Shanghai Synchrotron Radiation Facility (SSRF). Pressure was applied at room temperature and calibrated by using the ruby fluorescence shifts for all experiments [23].

## III. Results and Discussion

The crystal structure, phase purity, chemical composition and bandgap of the as-grown $Pd_3P_2S_8$ single crystals were examined by XRD, Raman, EDXS and optical transmittance measurements, for which the results are given by Fig. S1 and Fig. S2 in the Supplementary Materials [24]. Based on comparison with previous reports on $Pd_3P_2S_8$ single crystals, all the results consistently indicate a high-quality of the single crystals [24].

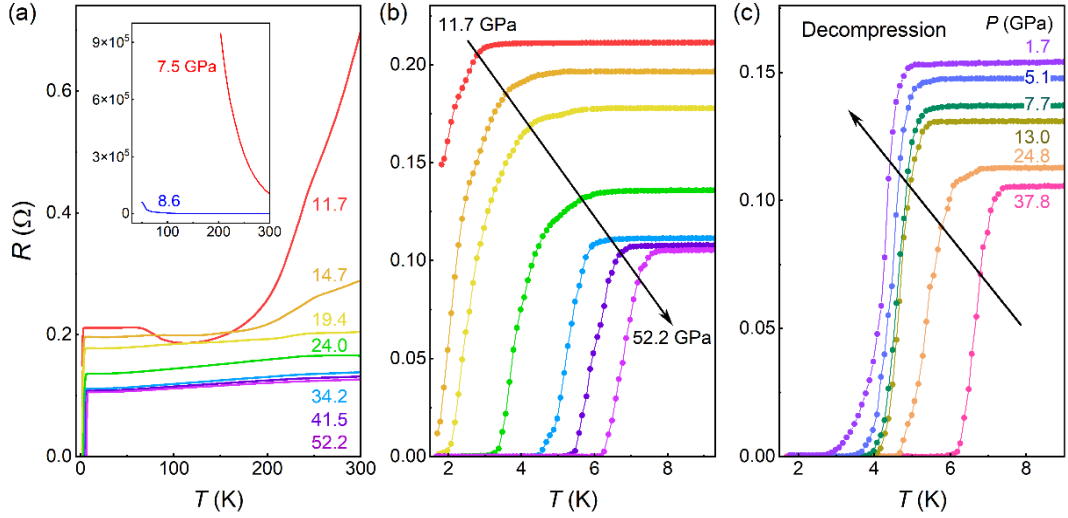

Figure 1 (color online). (a) Temperature-dependent resistance curves $R(T)$ for the $Pd_3P_2S_8$ single crystal measured at different pressures. The inset displays the curves for low pressures of 7.5 GPa and 8.6 GPa. (b) a zoomed-in view of the $R(T)$ curves near the superconducting critical temperatures. (c) the $R(T)$ curves upon decompression from 37.8 GPa to 1.7 GPa.

High-pressure resistance measurements were performed on a $Pd_3P_2S_8$ single crystal for investigating the pressure evolution of the electronic ground state. Due to the limited measuring range of instrument, resistance measurements are only available for pressures above 7.5 GPa, given that $Pd_3P_2S_8$ at ambient pressure is a semiconductor with large bandgap of 1.9-2.1 eV [20,21]. Figure 1 (a) presents the temperature-dependent resistance curves $R(T)$ for the single crystal measured at different pressures. At low pressures of 7.5-8.6 GPa, the $R(T)$ curves exhibit semiconducting behaviors. Meanwhile, the amplitude of resistance is dramatically suppressed with increasing pressure. From 11.7 GPa to above, however, a metallic behavior is instead observed. Upon entering the metallic phase, a sharp resistance drop is observed at an onset temperature of 2.6 K, as is shown in Fig. 1(b). When pressure is raised up to 19.4 GPa, zero resistance state is probed at 1.8 K, indicating the presence of superconductivity. Both the onset superconducting critical temperature $T_c^{onset}$ and zero-resistance temperature $T_c^{zero}$ display monotonous increases with increasing pressure to 52.2 GPa. Upon decompression [see Fig. 1(c)], we find that the superconductivity can be preserved to ambient condition though both $T_c^{onset}$ and $T_c^{zero}$ are slightly reduced.

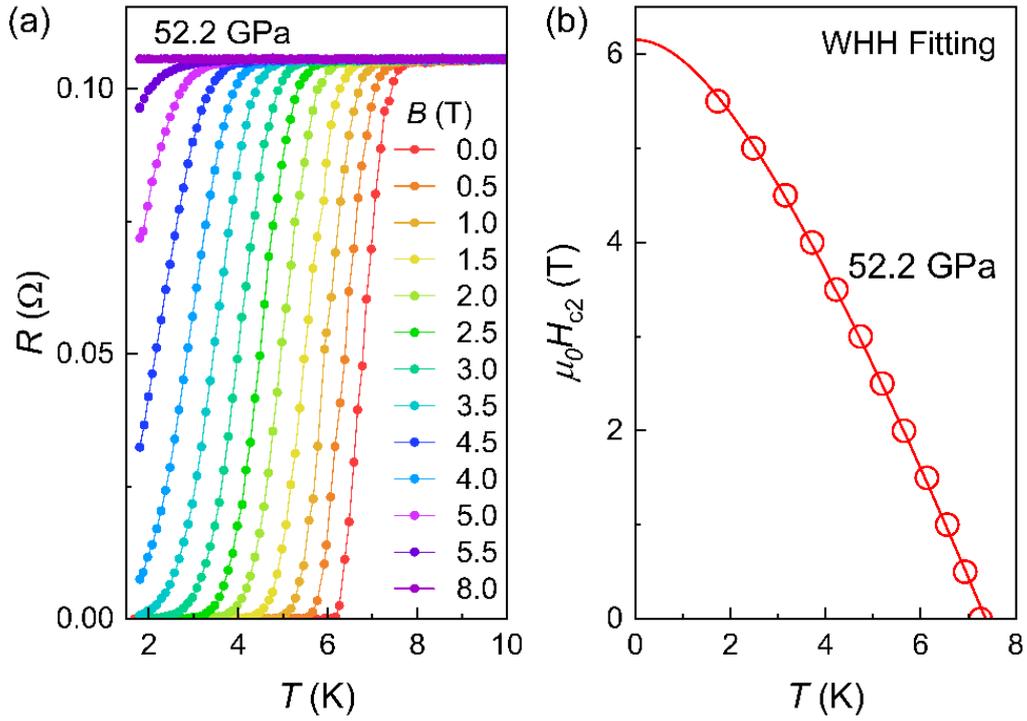

Figure 2 (color online). (a) Magnetic field-evolution of superconducting resistive transition measured at pressure 52.2 GPa. (b) The upper critical field $\mu_0H_{c2}(T)$ as a function of superconducting critical temperature $T_c^{90\%}$.

Figure 2 (a) presents the evolution of superconductivity against magnetic field which is measured at the highest pressure 52.2 GPa. As is expected, superconductivity is gradually suppressed with increasing magnetic field. In Fig. 2 (b) we summarize the extracted upper critical field $\mu_0H_{c2}$ as a function of critical temperature $T_c^{90\%}$, which is defined by the temperature where the normal-state resistance drops by 10%. By fitting the data to the Werthamer-Helfand-Hohenberg (WHH) model [26], we obtain a zero-temperature limit of the upper critical field $\mu_0H_{c2}(0)$ of 6.15 T.

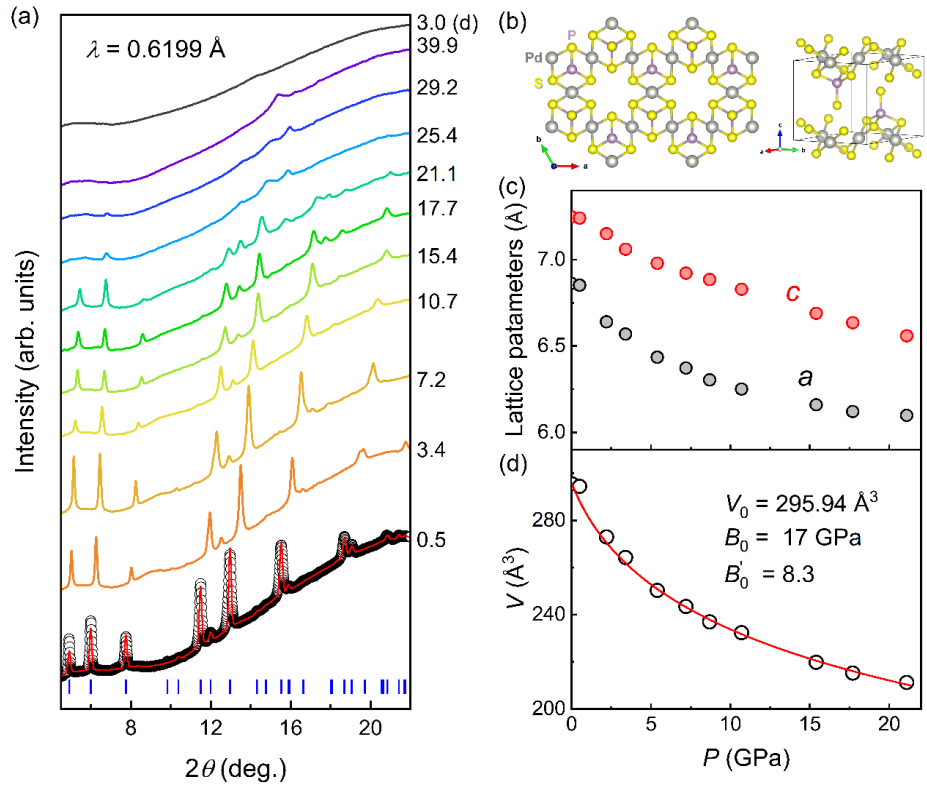

Figure 3 (color online). (a) High-pressure powder XRD patterns measured on grounded $Pd_3P_2S_8$ crystals for pressures up to 39.9 GPa, and then down to 3.0 GPa ("d" means decompression). For $P$ = 0.5 GPa, a red solid line presents a representative fitting. (b) Crystal structure of $Pd_3P_2S_8$ from a top view and side view, where the kagome lattice of Pd atoms is clearly shown. (c) Pressure dependences of the fitted lattice parameters $a$ and $c$. (d) Pressure dependence of the unit-cell volume $V$. The red solid line represents a fitting to the data using the third-order Birch-Murnaghan equation of states.

In order to find out the underlying structural evolution during the development of superconductivity, we performed angle-dispersive synchrotron XRD experiments on powdered crystals of $Pd_3P_2S_8$, for which the results are given in Fig. 3(a). For the initial pressure 0.5 GPa, all the diffraction peaks can be well indexed and fitted by the space group *P-3m1* (No. 164), which is the same as that at ambient pressure, as is schematically shown in Fig. 3 (b). Below 21.1 GPa, no new diffraction peak appears, indicating absence of any structural transition. Above 21.1 GPa, however, most of the diffraction peaks rapidly disappear and only very weak peaks can be seen at the highest pressures. These facts evidence that the system undergoes an amorphization at 21.1 GPa. When pressure is released to 3.0 GPa, none of the crystalline diffraction peaks is restored. The irreversible evolution of structure is consistent with the irreversible development of superconductivity as is presented by the transport measurements.

The lattice parameters *a*, *c*, and unit-cell volume *V* are extracted by fittings and plotted as functions of pressure in Fig. 3 (c) and (d). All these parameters monotonously decrease with pressure. Isothermal equations of state (EoS) were fitted on the unit-cell volume *V* by using the third-order Birch-Murnaghan formula [27]: $P = \frac{3}{2} B_0 [(\frac{V_0}{V})^{\frac{7}{3}} - (\frac{V_0}{V})^{\frac{5}{3}}][1 + \frac{3}{4}(B_0' - 4)[(V_0/V)^{\frac{2}{3}} - 1]]$, where $V_0$, $B_0$ and $B_0'$ are volume, bulk modulus $-V/(dV/dP)$, and first order derivative of the bulk modulus at ambient pressure, respectively. With $V_0$ fixed as 295.94 Å³, our fitting yields $B_0$ = 17 GPa, and $B_0'$ = 8.3.

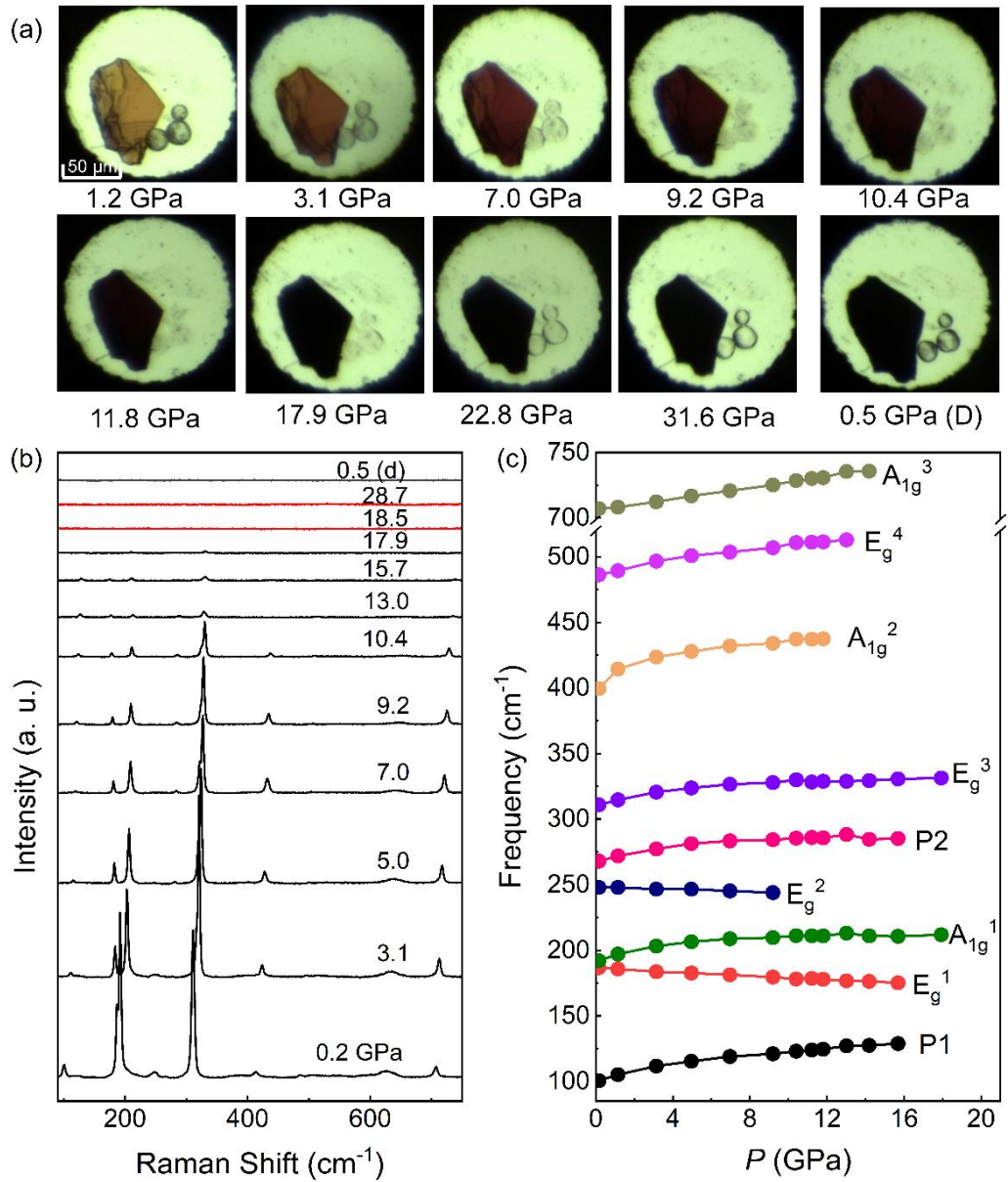

Figure 4 (color online). (a) Optical micrographs of the $Pd_3P_2S_8$ single crystal collected at different pressures. The last one "0.5 GPa (D)" is for a state after decompression. (b) Raman spectra of the $Pd_3P_2S_8$ single crystal for different pressures, which are collected at room-temperature with laser wavelength of $\lambda = 532$ nm. (c) Pressure-dependent frequencies of the detected Raman modes.

The high-pressure structure of the $Pd_3P_2S_8$ single crystal is further examined by Raman experiments. During the measurements, we incidentally collected the optical micrographs for the single crystal, which are shown in Fig. 4(a). From 1.2 GPa to 10.4 GPa, the single crystal is partially transparent and becomes darker and darker with increasing pressure, which corresponds to a gradually increased reflection of visible light due to the increment of charge carrier concentration. For pressures above 11.8 GPa, the sample becomes totally black, characterizing a metallic nature of the sample. After decompression to 0.5 GPa, the black color remains which agrees with the irreversible evolution of the transport properties.

Figure 4(b) presents the Raman spectra of the $Pd_3P_2S_8$ single crystal measured at different pressures. At 0.2 GPa, all the present Raman modes are consistent with those measured at ambient pressure [15,24,25]. With increasing pressure, the Raman spectral intensity exhibits a decline at 10.4 GPa and all Raman modes finally disappear above 17.9 GPa. When pressure is released down to 0.5 GPa, no Raman mode is stored, consistent with the irreversible evolution of XRD pattern.

The frequencies of the Raman modes are extracted and plotted as functions of pressure in Fig. 4(c). The smooth evolutions of the mode frequencies, in combination with the fact that no new Raman mode is present, evidence that no structural transition takes place below 17.9 GPa, which is in line with the XRD results. While most of the Raman modes expectedly show blueshifts with increasing pressure, only the modes $E_g^1$ and $E_g^2$ exhibit unusual redshifts. Generally, such kind of unusual redshifts can be related to structural instability [28-30]. According to previous theoretical and experimental studies, the Raman modes $E_g^1$ and $E_g^2$ in $Pd_3P_2S_8$ are respectively associated with the bending and tilting vibrations of the tetrahedral $[PS4]^{3-}$ units [15]. The redshifts of modes $E_g^1$ and $E_g^2$ here may imply a structural instability of the $[PS4]^{3-}$ tetrahedrons under pressure.

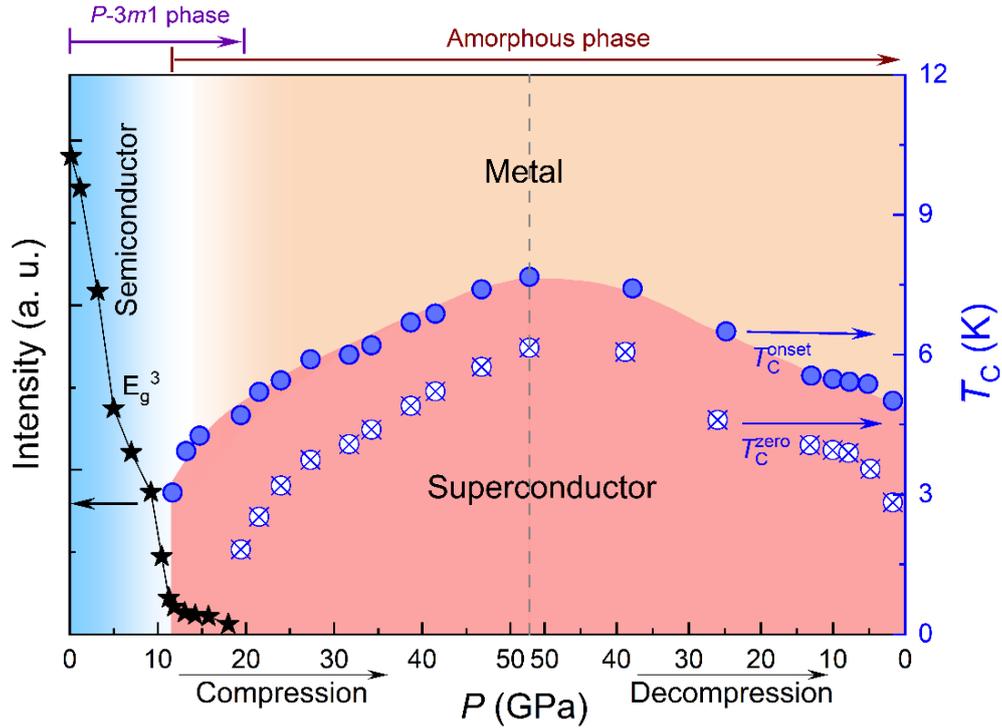

Figure 5 (color online). The phase diagram of $Pd_3P_2S_8$ as a function of pressure. The left axis stands for the Raman intensity of a representative mode $E_g^3$. The right axis corresponds to the superconducting critical temperatures $T_c^{onset}$ and $T_c^{zero}$. The colored areas are guides to the eyes, indicating the distinct conducting phases of semiconductor, metal, and superconductor.

All the results are summarized in a phase diagram in Fig. 5. The intensity of a representative Raman mode $E_g^3$ is also included. As is shown above, the XRD results evidence an amorphization of the system at ~20 GPa. Here the pressure dependence of the Raman intensity, which displays a sharp drop at ~11 GPa, indicates that the amorphous phase may already begin to form at this early pressure. Accompanied with the beginning of amorphization, metallization and onset superconducting transition are immediately observed. When pressure is raised to ~20 GPa where the amorphization completes, zero resistance state is detected and simultaneously the Raman intensity vanishes. Therefore, the emergence of superconductivity is attributed to the formation of amorphous phase. Moreover, XRD and Raman results consistently indicated that the crystal structure cannot be recovered after the pressure is released, which therefore explains the retainability of superconductivity upon decompression.

## IV. Conclusion

In summary, we have grown kagome single crystal $Pd_3P_2S_8$ and investigated the high-pressure electrical transport and structural properties. We find that a pressure of ~11 GPa induces a metallization and superconductivity in $Pd_3P_2S_8$. With increasing pressure, superconducting $T_c$ continuously increases. Structural analyses show that the occurrence of superconductivity is due to the formation of an amorphous phase. As a result of the irreversible nature of the amorphization, superconductivity can be preserved to ambient condition.

## Acknowledgements

This work was financially supported by the National Key Research and Development Program of China (Grant No. 2018YFA0305704), the National Natural Science Foundation of China (Grant Nos. 11874362, U1932152, 12174395, U19A2093, 12004004); the Natural Science Foundation of Anhui Province (Grant Nos. 2008085QA40, and 1908085QA18); the Key Project of Natural Scientific Research of Universities in Anhui Province (Grant Nos. KJ2021A0068, and KJ2021A0064); the Users with Excellence Project of Hefei Center CAS (Grant Nos. 2021HSC-UE008 and 2020HSC-UE015); and the Collaborative Innovation Program of Hefei Science Center, CAS (2020HSC-CIP014). A portion of this work was supported by the High Magnetic Field Laboratory of Anhui Province under Contract No. AHHM-FX-2020-02. Y.H. Zhou was supported by the Youth Innovation Promotion Association CAS (Grant No. 2020443). The X-ray diffraction experiment was performed at the beamline BL15U1, Shanghai Synchrotron Radiation Facility (SSRF).